\documentclass[
 reprint,
 amsmath,amssymb,superscriptaddress,
 aps,
 prl,
]{revtex4-1}

\usepackage{graphicx}
\usepackage{color}
\usepackage{dcolumn}
\usepackage{bm}

\begin{document}

\title{Anisotropic Fermi Contour of (001) GaAs Holes in Parallel Magnetic Fields}
\date{\today}

\author{D.\ Kamburov}
\author{M.\ Shayegan}
\affiliation{Department of Electrical Engineering, Princeton University, Princeton, New Jersey 08544, USA}
\author{R.\ Winkler}
\affiliation{Department of Physics, Northern Illinois University, DeKalb, Illinois 60115, USA}
\affiliation{Materials Science Division, Argonne National Laboratory, Argonne, Illinois 60439, USA}
\author{L.N.\ Pfeiffer}
\author{K.W.\ West}
\author{K.W.\ Baldwin}
\affiliation{Department of Electrical Engineering, Princeton University, Princeton, New Jersey 08544, USA}

\begin{abstract}
We report a severe, spin-dependent, Fermi contour anisotropy induced by parallel magnetic field in a high-mobility (001) GaAs two-dimensional hole system. Employing commensurability oscillations created by a unidirectional, surface-strain-induced, periodic potential modulation, we directly probe the anisotropy of the two spin subband Fermi contours. Their areas are obtained from the Fourier transform of the Shubnikov-de Haas oscillations. Our findings are in semi-quantitative agreement with the results of parameter-free calculations of the energy bands.
\end{abstract}

\pacs{}

\maketitle

The complex energy band structure of GaAs two-dimensional hole systems (2DHSs) has been the subject of continued research thanks to its fundamental importance and, more recently, for the potential application of 2DHSs in spintronics and quantum computing \cite{Hayden.PRL.1991, Heremans.SurfSci.1994, Brosh.PRB.1996, Lu.PRB.1999, Winkler.2003, Rokhinson.PRL.2004, Danneau.PRL.2006, Chesi.PRL.2011, Hirmer.PRL.2011}. The 2D hole dispersion is characterized by strong spin-orbit interaction, which leads to spin-splitting of the bands even in the absence of an external magnetic field and makes 2DHSs useful for spintronic devices \cite{Lu.PRB.1999, Winkler.2003, Rokhinson.PRL.2004}. In addition, the holes' wave functions have little overlap with the nuclei. The lack of overlap should significantly improve the spin coherence time, rendering the holes' spins promising candidates for quantum computing qubits \cite{Bulaev.PRL.2005, Heiss.PRB.2007, Gerardot.Nature.2008}.

Here we address the ability to manipulate the 2D holes' energy bands using a magnetic field ($B_\|$) applied parallel to the plane of a 2DHS and to directly probe the resulting distortions of the spin subband Fermi contours and the ballistic hole trajectories. The distortions are a result of the finite thickness of the (quasi-) 2D hole layer and the coupling of $B_\|$ to the holes' orbital motion. As we demonstrate, the Fermi contour distortion for the majority-spin holes is particularly significant and leads to a contour anisotropy of $ \sim 3:1$ for $B_\| \simeq 15$~T in our 175-\AA\--wide GaAs quantum well (QW) sample. This anisotropy is much larger than what is expected in 2D \textit{electron} systems confined to a similar GaAs QW \cite{Smrcka.1994}.

A direct and quantitative determination of the anisotropy of the 2D hole Fermi contours in a strong $B_\|$ is by itself of fundamental interest. Pioneering magneto-tunnelling measurements of the 2D hole energy band anisotropy, reported over twenty years ago, agreed surprisingly well with the results of simple ($4 \times 4$ Luttinger model) calculations of the bands at \textit{zero magnetic field} \cite{Hayden.PRL.1991}. This is very puzzling because the experimental data were taken at very high values of $B_\|$ (up to 25~T) and yet there was good agreement with zero-field dispersions. Subsequent theoretical calculations validated this puzzle as they showed that the agreement becomes worse if one uses more accurate energy band models (at zero magnetic field) \cite{Winkler.SurfSci.1994} or takes the large $B_\|$ into account \cite{Goldoni.PRB.1993}. The anisotropy is also relevant in measurements where magnetic focusing of ballistic holes is used for spatial spin separation \cite{Rokhinson.PRL.2004, Chesi.PRL.2011}. In such experiments a relatively strong $B_\|$ is often applied to partially spin-polarize the 2D holes. This $B_\|$ can cause a severe distortion of the hole Fermi contour. Our measurements of Fermi contour distortions and their close comparison with the results of state-of-the-art calculations therefore not only shed light on a long-standing problem, but they also have implications for the realization of devices whose operation depends on the ballistic transport of 2D holes.

\begin{figure*}[t]
\includegraphics[trim=0.6cm 0.8cm 0.4cm 1.0cm, clip=true, width=1\textwidth]{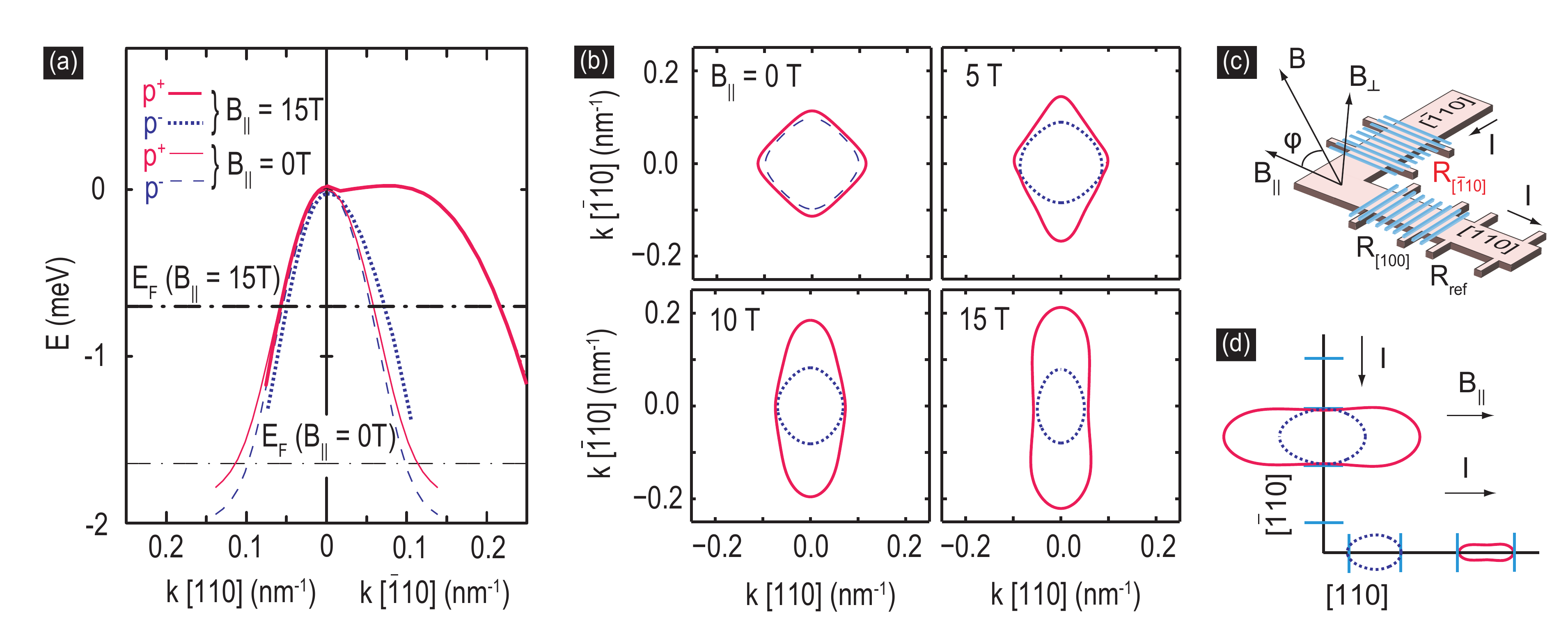}
\caption{\label{fig:Fig1}
(color online) (a) Self-consistently calculated dispersions for the majority-spin ($p^+$) and minority-spin ($p^-$) subbands along the $[110]$ and $[\overline{1}10]$ directions for a 2DHS of density $p=1.5 \times 10^{11}$ cm$^{-2}$ confined to a symmetric 175-\AA\--wide (001) GaAs QW. Thick solid and dotted curves represent the dispersions at $B_\|=15$~T applied along $[110]$, while thin solid and dashed curves are for $B_{||}=0$. (b) The calculated 2D hole Fermi contours for the $p^+$ and $p^-$ subbands in a parallel magnetic field applied along the $[110]$ direction are given in solid red and dashed/dotted blue, respectively. (c) Schematic of the experimental set-up, indicating the orientation of the Hall bar arms and the applied magnetic field. (d) The geometry of the Hall bar is designed to use the commensurability of the ballistic cyclotron orbits in real space with the period of the potential modulation induced by the stripes to probe the size of the Fermi wave vector $k_F$ along the $[110]$ and $[\overline{1}10]$ directions directly (see text).}
\end{figure*}

Figure~\ref{fig:Fig1} highlights the key points of our study. In Fig.~\ref{fig:Fig1}(a) we show the results of a parameter-free calculation of the 2DHS dispersions, based on an $8 \times 8$ Kane Hamiltonian that takes into account the spin-orbit interaction and the nonparabolicity of the 2D hole bands in our sample \cite{Winkler.2003}. The Fermi contours are given in Fig.~\ref{fig:Fig1}(b). At $B_\| = 0$~T, the Fermi contours of the two spin-split bands differ slightly from each other. With the application of $B_\|$ along the $[110]$ direction, the two contours change dramatically. The majority-spin ($p^+$) and the minority-spin ($p^-$) Fermi contours both become elongated along the direction perpendicular to $B_\|$. This anisotropy is very different for the two spin subbands, being $\sim 3:1$ ($\sim 1.6:1$) in the majority (minority) spin subband \cite{Smrcka.1994}. The pure Zeeman spin splitting of the hole states at $k=0$ is rather small, but the spin splitting shows a pronounced $\bm{k}$ dependence, which reflects the interplay of spin-orbit coupling and heavy hole-light hole coupling \cite{Winkler.2003}. Holes are also transferred from the $p^-$ to the $p^+$ contour with increasing $B_\|$ as evidenced by the enhanced area of the $p^+$ contour. Note that the real-space hole trajectories (see Fig.~\ref{fig:Fig1}(d)) are rotated by $90^{\circ}$ with respect to the Fermi contours \cite{Ashcroft.1976} so that, as expected for a quasi-2D carrier system with finite layer thickness, the real-space trajectories are longer along the direction of $B_\|$ ($[110]$) and are squeezed perpendicular to $B_\|$ ([$\overline{1}10$]) \cite{dresselhaus}.

In our study we employ surface-strain-induced commensurability oscillations (COs) \cite{Skuras.APL.1997, Endo.PRB.2000, Endo.PRB.2001, Endo.PRB.2005, Kamburov.PRB.2012}, triggered by a periodic superlattice, to directly map the Fermi wave vectors in two perpendicular directions, $[\overline{1}10]$ and $[110]$, as shown in Figs.~\ref{fig:Fig1}(b) and (c) \cite{footnote}. The magneto-resistance of such samples exhibits minima at the electrostatic commensurability condition $2R_C/a=i-1/4,$ where $i=1,2,3 \ldots$ \cite{Endo.PRB.2000, Endo.PRB.2000, Endo.PRB.2001, Endo.PRB.2005, Kamburov.PRB.2012, footnote, Weiss.EL.1989, Winkler.PRL.1989, Gerhardts.PRL.1989, Beenakker.PRL.1989, Beton.PRB.1990, Peeters.PRB.1992, Mirlin.1998, Gunawan.2004}. Here $2R_C = 2k_F/eB$ is the cyclotron diameter along the modulation direction and $a$ is the period of the potential modulation ($k_F$ is the Fermi wave vector \textit{perpendicular} to the modulation direction) \cite{Gunawan.2004}. The anisotropy of the cyclotron diameter and/or the Fermi contour can therefore be directly determined from COs measured along the two perpendicular arms of an L-shaped Hall bar as shown in Figs.~\ref{fig:Fig1}(c) and (d). The COs for the arms along $[110]$ and $[\overline{1}10]$ yield $k_F$ along $[\overline{1}10]$ and $[110]$, respectively. In a semiclassical picture, COs can be understood similar to Shubnikov-de Haas (SdH) oscillations \cite{Beenakker.PRL.1989, Ashcroft.1976} which show that, within the range of validity of the semiclassical approximation \cite{Winkler.2000}, COs yield the (extremal) Fermi wave vector perpendicular to the modulation direction, irrespective of details of the dispersion such as nonparabolicity and anisotropy. In our measurements, we also recorded SdH oscillations in an unpatterned part of the sample to probe the area enclosed by each of the Fermi contours.

We prepared strain-induced superlattice samples with a lattice period of $a=175$ nm from a 2DHS confined to a 175-\AA\--wide GaAs QW grown via molecular beam epitaxy on a (001) GaAs substrate. The superlattice is made of negative electron-beam resist and modulates the 2DHS potential through the piezoelectric effect in GaAs \cite{Kamburov.PRB.2012}. The QW, located 131 nm under the surface, is flanked on each side by 95-nm-thick Al$_{0.24}$Ga$_{0.76}$As spacer layers and C $\delta$-doped layers. The 2DHS density at $T\simeq$ 0.3 K is $p\simeq 1.5 \times 10^{11}$ cm$^{-2}$, and the mobility is $\mu=1.2\times10^{6}$ cm$^2$/Vs. The sample has two Hall bars, oriented along the [$110$] and [$\overline{1}10$] directions, as schematically illustrated in Fig.~\ref{fig:Fig1}(c). Current was passed along the two Hall bar arms and the longitudinal resistances along the arms were measured simultaneously. We made measurements by first applying a fixed, large magnetic field in the plane of the sample along $[110]$. The sample was then slowly rotated to introduce a small magnetic field ($B_{\perp}$) perpendicular to the 2DHS; this $B_{\perp}$ is what induces COs or SdH oscillations in our sample \cite{Tutuc.PRL.2001}. The magnitude of $B_{\perp}$ was extracted from the Hall resistance which we measured in an unpatterned region of the sample along with the resistances of the two patterned regions. Note that, when the applied field $B$ is large compared to $B_{\perp}$, the parallel component of the field, $B_\|=\sqrt{B^2-B_{\perp}^2}$, remains essentially fixed and equal to $B$ as we rotate the sample and take data \cite{Tutuc.PRL.2001}. Also, we tilted the sample around the [$\overline{1}10$] direction so that $B_\|$ was always along [$110$]. We performed the experiment using low-frequency lock-in techniques in a $^3$He refrigerator with a base temperature of $T\simeq$ 0.3 K.

\begin{figure}[!h]
\includegraphics[trim=0.2cm 0.30cm 0.05cm 0.2cm, clip=true, width=0.48\textwidth]{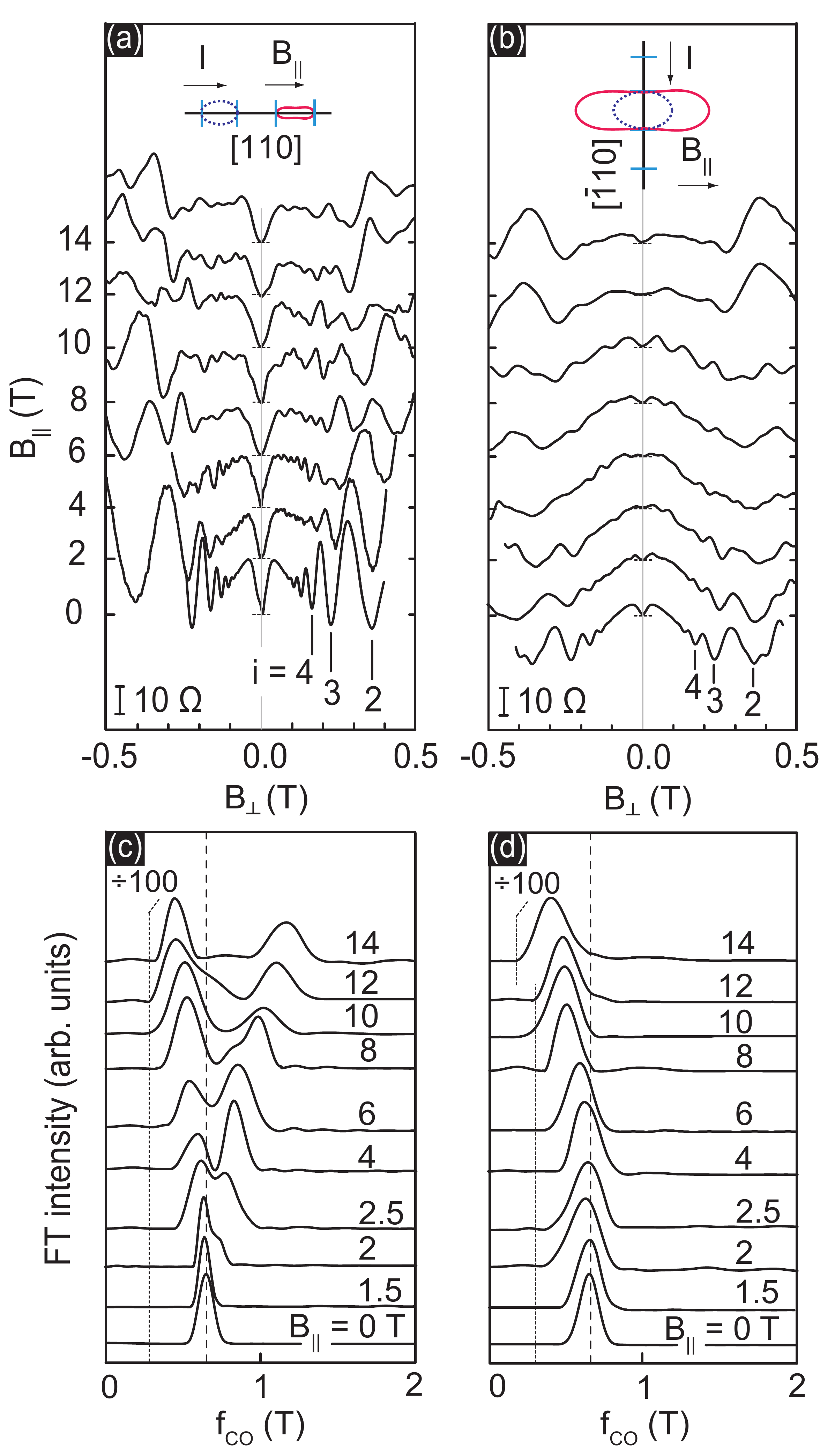}
\caption{\label{fig:Fig2}
(color online) (a), (b) Magnetoresistance data measured across the patterned sections of the L-shaped Hall bar in the $[110]$ and $[\overline{1}10]$ directions at different values of $B_\|$. Each trace is vertically offset so that its resistance value at $B_{\perp}=0$, marked by a horizontal bar, lines up with the value of $B_\|$ at which the trace was taken (shown as the y-axis). The expected positions $i=2,3,4$ of the COs minima for the bottom ($B_\|=0$) traces are indicated with vertical lines. (c), (d) Normalized Fourier transform spectra of the COs data shown in (a) and (b), respectively. The $B_\|=0$ anticipated COs frequency, based on a spin-degenerate, circular Fermi contour, is marked with dashed lines. The low-frequency parts of the spectra (below the vertical dotted lines) are severely affected by the Hamming window used in the Fourier analysis and are shown here suppressed by a factor of 100.}
\end{figure}

The magnetoresistance data from the two perpendicular Hall bar arms are shown in Figs.~\ref{fig:Fig2}(a) and (b). In each figure the bottom trace, which was taken in the absence of $B_\|$, exhibits clear COs. The positions of the resistance minima agree well with those predicted by the commensurability condition for a 2DHS with a circular, spin-degenerate, Fermi contour; the latter are indicated with indexed vertical lines \cite{Kamburov.PRB.2012,resolution}. The Fourier transforms (FTs) of these two traces are shown as the bottom curves in Figs.~\ref{fig:Fig2}(c) and (d). Each of the two FT spectra exhibits one peak whose position ($\simeq 0.64$~T) agrees with the frequency $f_{CO}=2 \hbar k_F/ea=0.64$~T expected for a circular, spin-degenerate Fermi contour with $k_F= \sqrt{2 \pi p}$ \cite{Kamburov.PRB.2012, resolution}. For sufficiently large values of $B_\|$ ($\geq 2$~T), the peak in the FTs for the [110] Hall bar data (Fig.~\ref{fig:Fig2}(c)) splits into two peaks, and the splitting increases with increasing $B_\|$. In sharp contrast to this behavior, the $[\overline{1}10]$ Hall bar data (Fig.~\ref{fig:Fig2}(d)) show only one peak whose position moves to smaller frequencies as $B_\|$ increases.

\begin{figure}[t]
\includegraphics[trim=0cm 0.2cm 0cm 0cm, clip=true, width=0.48\textwidth]{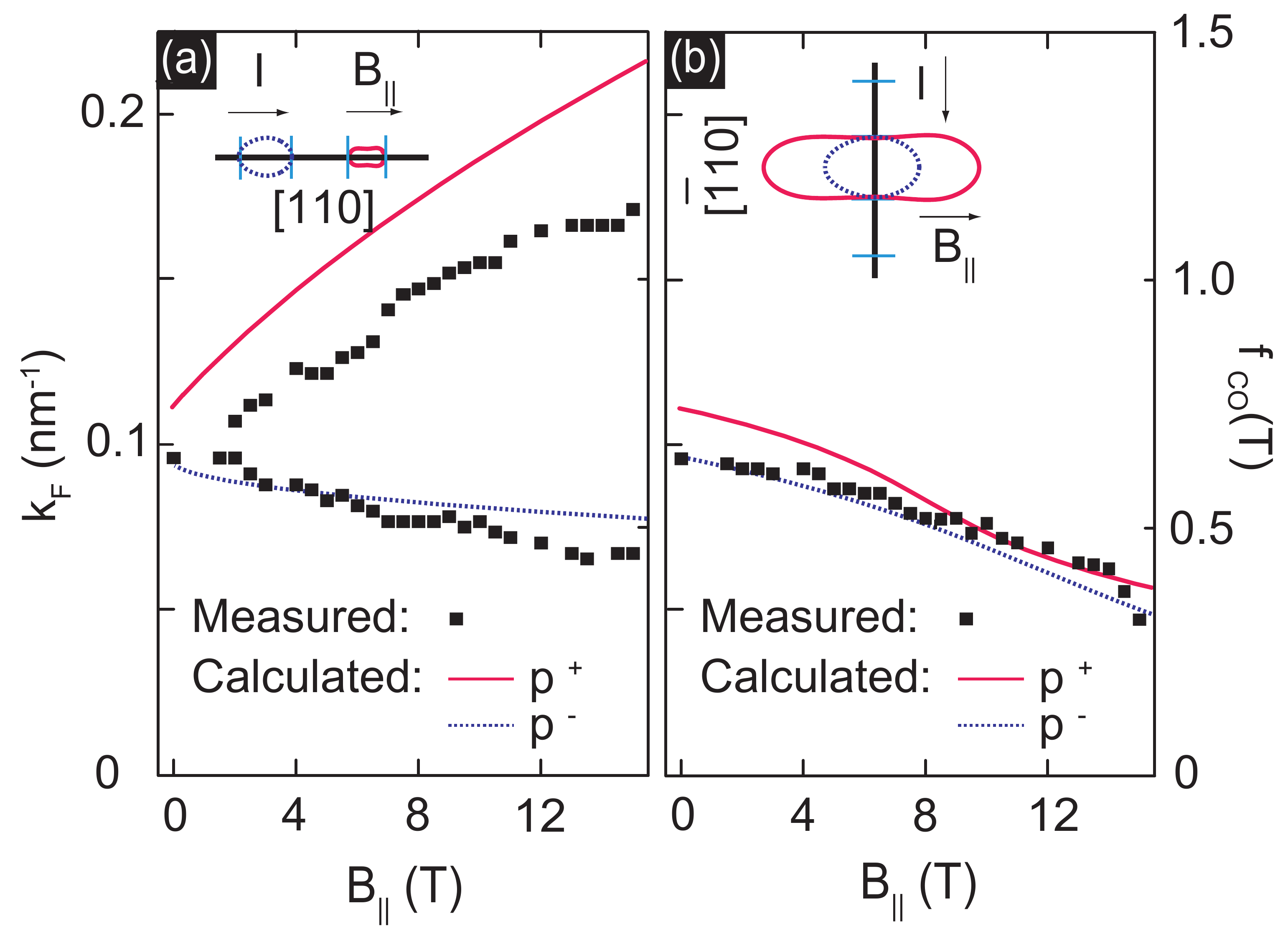}
\caption{\label{fig:Fig3}
(color online) (a), (b) Summary of the peak positions of the COs Fourier spectra for the two Hall bar arms. The left axis shows the deduced Fermi wave vectors $k_F$ according to $k_F=eaf_{CO}/2 \hbar$. The experimental data are shown by square symbols. The lines represent the corresponding calculated values, based on $k_F$ of the $p^+$ and $p^-$ contours. }
\end{figure}

Figure~\ref{fig:Fig3} summarizes the measured $f_{CO}$ as a function of $B_\|$ and the corresponding deduced Fermi wave vectors $k_F$ (left axis). In this figure, we also plot the values of $k_F$ as predicted by our calculations of the Fermi contour shapes. The qualitative agreement between the measured and calculated $k_F$ in Fig.~\ref{fig:Fig3} is clear. The agreement is in fact quantitatively good except for the $p^+$ Fermi contour along $[\overline{1}10]$ where the elongation deduced from the experimental data is smaller than expected from the calculations (Fig.\;\ref{fig:Fig3}(a)). We do not know the source of this disagreement at the moment. Despite this discrepancy, however, the overall agreement between the measured and calculated $k_F$ is remarkable, considering that there are no adjustable parameters in the calculations. The results of Fig.~\ref{fig:Fig3} clearly point to a severe spin-dependent distortion of the Fermi contours and the associated real-space ballistic hole trajectories in the presence of a moderately strong $B_\|$.

\begin{figure}[!h]
\includegraphics[trim=0.1cm 0.4cm 0cm 0.2cm, clip=true, width=0.48\textwidth]{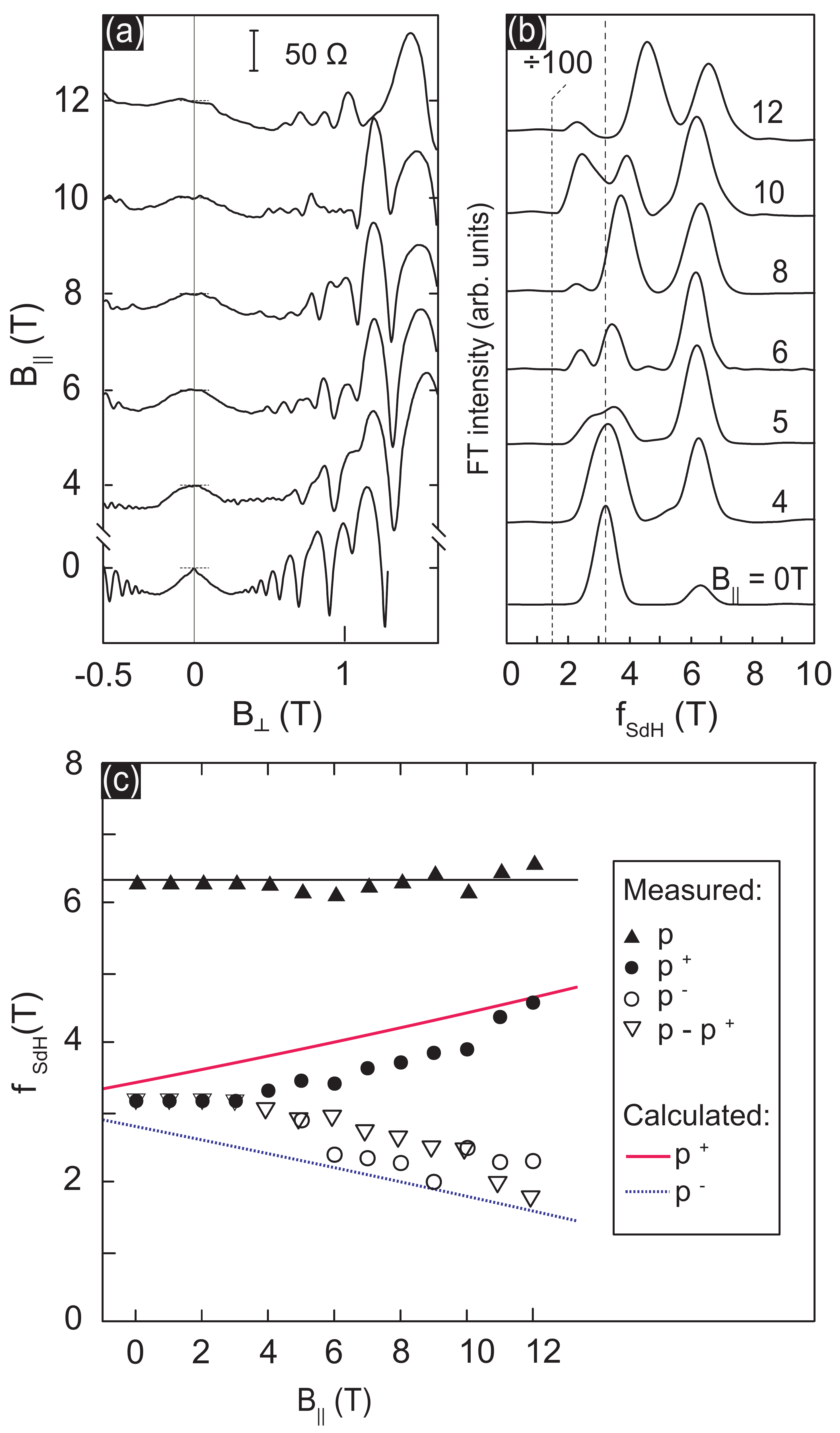}
\caption{\label{fig:Fig4}
(color online) (a) Shubnikov-de Haas oscillations measured in the reference (unpatterned) region of the Hall bar as $B_\|$ increases. Similar to Figs.~\ref{fig:Fig2}(a) and (b), the traces are vertically offset for clarity. (b) Fourier transform spectra of the SdH oscillations as a function of $B_\|$. The dashed line shows the expected position of the $B_\|=0$ spin-unresolved FT peak. The signal in the region to the left of the vertical dotted line is shown suppressed. (c) Summary of the measured (symbols) and calculated (lines) SdH frequencies. }
\end{figure}

The COs data we present in Figs.~\ref{fig:Fig2} and \ref{fig:Fig3} probe the size of the 2DHS Fermi contours in two specific directions in $k$-space. For completeness, we also probed the areas of the Fermi contours by measuring SdH oscillations in an unpatterned region of the sample. Figure~\ref{fig:Fig4}(a) shows the magnetoresistance traces at different values of $B_\|$. The FTs of these traces are shown in Fig.~\ref{fig:Fig4}(b). For $B_\|=0$ and also at low values of $B_\|$, we observe two peaks. The position of the peak at 3.1~T matches the value of $(h/2e)p \simeq 3.1$~T expected for spin-unresolved SdH oscillations of holes of density $p \simeq 1.5 \times 10^{11}$ cm$^{-2}$, and the peak at 6.2~T corresponds to spin-resolved oscillations. Starting at $B_\| \simeq 5$~T, the spin-unresolved peak at 3.1~T begins to split, with the upper and lower peaks corresponding to the areas (hole densities) of the $p^+$ and $p^-$ subbands, respectively.

In Fig.~\ref{fig:Fig4}(c) we plot, as a function of $B_\|$, a summary of the three measured SdH frequencies ($f_{SdH}$), corresponding to $p^+$, $p^-$, and the total density, $p$. We also plot (open triangles) the difference between the measured frequencies $p$ and $p^+$ as an alternative measure of $p^-$. To compare the experimental data with the results of our energy band calculations, in Fig.~\ref{fig:Fig4}(c) we show two curves corresponding to the areas (divided by $e/h$) of the calculated Fermi contours (see contours shown in Fig.~\ref{fig:Fig1}(a)). There is overall good agreement between the measured and calculated Fermi contour areas in Fig.~\ref{fig:Fig4}(c), although the measured splitting between the $p^+$ and $p^-$ bands is somewhat smaller than the calculations predict. A similar discrepancy has been reported before indicating that the SdH oscillations may not be simply related to the zero-magnetic-field hole densities \cite{Winkler.2000}. This precludes us from making a direct comparison between the CO and SdH data.

Our results presented here demonstrate the tuning of the GaAs 2D hole dispersion anisotropy through the application of an in-plane magnetic field. We provide data which directly probe the anisotropy and the size of the Fermi contours. The experimental data are in semi-quantitative agreement with the results of a parameter-free energy band model based on the $8 \times 8$ Kane Hamiltonian. We find a severe spin-dependent anisotropy of the 2D hole Fermi contours stemming from the combined effect of the strong $B_\|$ coupling to the orbital motion, the large spin-orbit interaction in the GaAs valence band and heavy hole-light-hole coupling \cite{Winkler.2003}. We emphasize that the anisotropy in our hole sample is much larger than our calculations predict for quasi-2D \textit{electrons} confined to a similar QW \cite{Smrcka.1994}.

\begin{acknowledgments}
We acknowledge support through the DOE BES (DE-FG02-00-ER45841) for measurements, and the Moore Foundation and the NSF (ECCS-1001719, DMR-0904117, and MRSEC DMR-0819860) for sample fabrication and characterization. Work at Argonne was supported by DOE BES under Contract No.\ DE-AC02-06CH11357. We thank Tokoyama Corporation for supplying the negative e-beam resist TEBN-1 used to make the samples.
\end{acknowledgments}


\begin{thebibliography}{99}
\bibitem{dresselhaus} Our calculations for a symmetric QW fully take into account Dresselhaus spin-orbit coupling. In the presence of an in-plane magnetic field in [110] direction the dispersions for $\bm{k}$ in $\pm [\overline{1}10]$ direction are thus slightly different; see Fig.~\ref{fig:Fig1}(b). The difference between these branches is too small to be resolved experimentally. In Fig.~\ref{fig:Fig1}(a) we show for $B_\| = 15$~T the dispersion in $[\overline{1}10]$ direction. Note also that the dispersion curves $E(\bm{k})$ for $B_\| = 0$~T and $B_\| = 15$~T were shifted vertically by different amounts to coincide at $\bm{k} = 0$.
\end{thebibliography}
\end{document}